# Segmental Dynamics in the Strain-hardening Regime for Poly(methyl methacrylate) Glasses with and without Melt-stretching

AUTHOR NAMES

*Enran Xing, Trevor Bennin, Masoud Razavi, M. D. Ediger**

AUTHOR ADDRESS

Department of Chemistry, University of Wisconsin – Madison, Madison, Wisconsin 53706, United States

*Corresponding Author (ediger@chem.wisc.edu)

ABSTRACT

Strain-hardening is a feature of polymer glasses during large deformation, which helps to stabilize the glasses against breakage. Experimentally, little is known about the segmental dynamics during strain-hardening, and such data is important for building a molecular-level theory of polymer glasses deformed in this regime. Here, using a photobleaching technique, we measured the segmental dynamics of lightly-crosslinked poly(methyl methacrylate) (PMMA) glasses with and without melt-stretching, which were deformed into the strain-hardening regime with local engineering strain rates from $10^{-4.6}s^{-1}$ to $10^{-4}s^{-1}$ at $T_g$-23K and $T_g$-33K. We find that melt-stretched PMMA glasses show a more prominent strain-hardening feature and faster segmental dynamics by a factor of about 15% compared to PMMA without melt-stretching. At a given true strain rate, the segmental dynamics of PMMA without melt-stretching are accelerated in the deep strain-hardening regime from the value just beyond yield, by up to 40% at 0.8 true strain. Our observations are in agreement with previously published simulation results.

## INTRODUCTION

Yielding and strain-hardening are two important features for plastic deformation of polymer glasses. Both features serve to stabilize the material against breakage: while yielding prevents brittle breakage at a few percent of strain[1-2], strain-hardening acts against strain localization and enables the sample to be deformed to a strain of 100% or more[3]. Considerable efforts have been made to obtain a microscopic understanding of these two processes. For yielding, previous work has brought out important insights[4-14]. Yielding was found to be closely related to the large acceleration of segmental dynamics from the undeformed state to the yield point[4-8], and this transition is interpreted as a process in which the deformation accelerates the rate at which segments escape from their local cages. Moreover, during a strain-rate controlled deformation, a power law relationship was found between the segmental relaxation time and the strain rate of a polymer glass in the flow state[15-16], with an exponent close to -1. On the other hand, our knowledge about strain-hardening from a microscopic perspective is more limited.

The understanding of strain-hardening has evolved a lot in the past decades. Early models[17-20] related strain-hardening to rubber elasticity that arises from the entangled chain network within a polymer glass, and the increase of stress during strain-hardening was viewed to originate from the chains' tendency to maximize the conformational entropy[21]. But this view was inconsistent with subsequent experimental results[22-23]. First, the fitted entanglement density from strain-hardening modulus could be two orders of magnitudes larger than that derived from the rubbery plateau modulus[22]. Second, the strain-hardening modulus decreases as the temperature increases, in contrast to the prediction for stress of an entropic nature[23]. In recent years, simulations by Hoy and Robbins[24-26] found that strain-hardening is a kinetic effect, rather than a thermodynamic effect; strain-hardening scales with flow stress[24] and that the stress increase during strain-hardening is highly correlated with the increase of local particle rearrangement rates[25-26]. More specifically, the stress derived from the dissipated heat, which contributes more than 80% of the total stress in the moderate strain-hardening regime, almost exactly follows the evolution pattern of the plastic rearrangement rate. These results suggest that, in addition to the close correlation with yielding, segmental dynamics might also play an important role during strain-hardening.

However, simulations and theory have opposite findings for the evolution of segmental dynamics during a constant true strain rate deformation. Almost all simulations suggested that the segmental dynamics accelerate under such a deformation[25-28], reaching a factor of 2 at a true strain of 100%[27]. While in an early version of the nonlinear Langevin equation (NLE) theory[29], strain-hardening was attributed to the suppression of local density fluctuations, an effect that slows down the segmental dynamics. Direct measurements of segmental dynamics are needed to address this controversy. In addition, it has been shown that melt-stretching followed by quenching can transform brittle polymer glasses into ductile glasses[30-32]. As a result of this process, strain-hardening is more prominent and occurs at smaller strain. Comparing the segmental dynamics between samples with and without melt-stretching allows a test of how strain-hardening is correlated with the segmental dynamics while other parameters, like strain rate and strain, are controlled.

Here, we investigate the change of segmental dynamics during strain-hardening with a photobleaching technique[4, 33-34], which measures the probe reorientation time of fluorescent probe molecules dispersed inside a polymer glass, a quantity that has been shown to be closely related to the segmental dynamics. Earlier work using the photobleaching technique has provided important insights for polymer glasses under creep deformations[33-37] or constant strain rate deformations[4, 15-16] just beyond yield. However, due to sample breakage issues, systematic experiments on samples deformed deep into the strain-hardening regime under strain-rate-controlled deformations haven't been reported previously. In the current work, making use of an upgraded sample cell that can quench the sample with cold nitrogen gas and generate polymer glasses with high ductility and enhanced strain-hardening, we performed photobleaching measurements on lightly-crosslinked poly(methyl methacrylate) (PMMA) glasses with and without melt-stretching, which were deformed into the strain-hardening regime with local engineering strain rates ranging from $10^{-4.6}s^{-1}$ to $10^{-4}s^{-1}$ at $T_g$-23K and $T_g$-33K.

We find that, in agreement with simulation results[25, 27], the segmental dynamics accelerate when more prominent strain-hardening occurs. Melt-stretched PMMA glasses, which show stronger strain-hardening than samples without melt-stretching, also have segmental dynamics that are about 0.05 decade faster. For quenched samples without melt-stretching, under a given true strain rate, the segmental dynamics were accelerated from the value just beyond yield by about 0.15 decade when a true strain of 0.8 was reached. These results will be useful for building a molecular-level theory of polymer glasses deformed in this regime.

## METHODS

**Sample preparation.** The fabrication protocol of the lightly-crosslinked poly(methyl methacrylate) (PMMA) samples used in the current work has been described previously[4, 16]. Briefly, the stock solution with 98.5 wt% methyl methacrylate (MMA), 1.5 wt% ethylene glycol dimethacrylate (EGDMA) and ~$5\times10^{-6}$ M optical probes, N,N′-dipentyl-3,4,9,10-perylenedicarboximide (DPPC), was mixed with the polymerization initiator, benzyl peroxide (~0.1 wt%), and put in a 345K water bath for pre-polymerization. When the mixture became viscous enough, it was transferred to molds made of two 2 × 3 in. microscope slides, with aluminum foil as spacers and clamped by binder clips on both ends. Then, the molds were kept at 345K under nitrogen for 24h. After that, the PMMA films were removed by sonication and cut into "dog bone" shaped samples with a custom-made die cutter; the active deformation region had a length of 13 mm and a width of 2mm. (The length is smaller than that in our previous work[4], to allow the melt-stretched sample to fit into our temperature-controlled cell.) Before use, the samples were kept at 415K for 24h under nitrogen to ensure the completion of the polymerization. The samples have curved thickness profiles, which are the thinnest in the middle with thicknesses ranging from 25-55 μm. The glass transition temperature, $T_g$, of the samples is 403±1K, as determined by the midpoint of the transition of the second DSC heating scan with a rate of 10K/min.

**Quenching apparatus.** All samples in the current study were quenched to the testing temperatures from above $T_g$ with cold nitrogen gas using a custom-built quenching apparatus. Nitrogen gas passed through a coil emersed in liquid nitrogen, and its temperature was monitored afterwards. When the gas reached around 140K, it was directed into the sample cell. Due to the fluctuations of the flow rate and the temperature of the nitrogen gas, thermal histories between different experiments varied slightly; the times for the sample cell to reach the testing temperatures varied between 2-4 min and there were undershoots of the cell temperature by about 5K (the undershoot of the sample temperature is likely larger). However, such variations had negligible impact on the segmental dynamics of the samples beyond yield, as confirmed by reproducible results from experiments with identical control parameters.

In the quenching process described above, the cell has a cooling rate of about 25K/min. We estimate the cooling rate of the sample to be about 100K/min, based on the stress behavior of melt-stretched samples, which indirectly indicates the time when the sample enters the glassy state from above $T_g$. Specifically, after melt-stretching at 420K($T_g$+17K), the stress relaxation was reversed (due to thermal contraction) about 10s after the quenching started. We interpret this to indicate that it takes about 10s for the sample's temperature to decrease by 17K. In comparison to the slow-cooling protocol with a rate of 1K/min, which was used in earlier works, the quenching protocol shortened the relaxation time measured in the glassy state. If we compare samples with the two cooling rates after ~20 minutes at the testing temperature (380K), the quenched (and also the melt-stretched) sample had a relaxation time about 1.2 decades faster (370s vs. 5700s) than a slow-cooled sample.

**Thermal and mechanical protocol.** The deformation apparatus has been introduced in a previous publication[4]. In brief, the sample was held by two clips, one of which was movable during deformation while the other one was not. And then the sample was placed into a brass cell. Tensile deformation with constant speed was driven by a linear actuator, which was connected to the moving clip through a U-shaped arm. The force is monitored by a load cell located between the sample and the linear actuator.

The samples for the experiments reported here were prepared with one of two protocols, either quenching alone or melt-stretching followed by quenching. These two types of samples will be called “quenched sample” and “melt-stretched sample” in this paper.

Quenched samples were annealed at 420K($T_g$+17K) for at least 25 min before being quenched to the testing temperatures (380K or 370K). The thermal protocol for melt-stretched samples was the same as quenched samples except that before the quenching step, the samples in the melt state (420K) were stretched at 0.5mm/s until reaching a stretching ratio ($\lambda$) of 2. Note that although the quenching started immediately after melt-stretching, given the quenching rate (~100K/min) that the apparatus achieves, segmental relaxation would occur before the sample is vitrified. After melt-stretching, the stretching ratio at the local measurement area (see below) has been confirmed to be equal to the global value, $\lambda$=2, which indicates that the deformation was homogeneous in the melt state.

For both quenched and melt-stretched samples, the mechanical deformation started 29±2min after the sample temperature was stabilized below $T_g$ at the testing temperatures. For quenched samples, the sample length, $L_0$, is 13mm. For melt-stretched samples, $L_0$ is 26mm. Global strain rates for glassy deformations are calculated based upon these values of $L_0$. Both types of samples were deformed with constant global strain rates of $2\sim5\times10^{-5}$ /s. We confirmed that the melt-stretching and the deformation in the glassy state did not permanently alter the samples: after deformation in the glassy state, all samples were able to return to their original states after annealing above $T_g$, as indicated by the completely recovered sample lengths and the reproducible mechanical and optical data in subsequent deformations.

Because of the inhomogeneous thickness profile, strain localization happens during glassy deformation. This occurred after yield near the thinnest part of a sample; the exact position was determined from a preliminary deformation with a final global strain of about 15%. All optical measurements were performed in this strain localization area. Local deformation information was collected by photobleaching lines parallel and perpendicular to the deformation axis (x-axis) and by taking pictures of this pattern over the entirety of a deformation, a procedure described in detail previously[34]. In brief, the change of the spacing between lines perpendicular to the deformation axis was used to calculate the local strain; while the contraction ratio, R, along the y-axis was determined from the spacing change of the lines parallel to the x-axis; the contraction along the y-axis and z-axis was assumed to be identical. Below the local strain will be presented as either "engineering local strain" or "true local strain". The former was determined as $\varepsilon_{eng}=(L(t)-L(0))/L(0)$ and the later as $\varepsilon_{true}=\ln(L(t)/L(0))$, in which L(t) is the spacing between bleaching lines perpendicular to the deformation axis at a certain time point and L(0) is the initial spacing (after any melt-stretching but prior to glassy deformation). Local strain rates were attained by taking time derivatives of the polynomial fittings of the local strain profiles. The contraction ratio along the y- and z- axes was used to calculate the true stress, which takes the shrinkage of the cross-section area during deformation into account, $\sigma_{true} = F/(A_0 \cdot R^2)$, in which F is the force, $A_0$ is the initial cross-section area and R is the contraction ratio.

**Photobleaching technique.** The photobleaching technique measures the reorientation time of fluorescent probes (DPPC) inside a polymer glass[4, 34], which has been confirmed to be a good reporter of segmental dynamics[38]. At the beginning of such a measurement, a linearly polarized laser beam (wavelength= 532nm) is used to preferentially photobleach probe molecules whose transition dipoles are aligned with the polarization state of the beam. Then, the photobleached area is exposed to a weak circularly polarized laser beam and the fluorescent light from the unbleached population of the probe molecules are collected and separated into two channels with polarization states parallel and perpendicular to that of the bleaching beam. The intensity of the perpendicular channel, $I_{\perp,b}$, is higher than that of the parallel channel, $I_{\parallel,b}$. The differences between the fluorescence intensity of the bleached area ($I_{\perp,b}$ or $I_{\parallel,b}$) and the neighboring unbleached area ($I_{\perp,un}$ or $I_{\parallel,un}$) are used to calculate the optical anisotropy, r(t)[34]. And the decay of r(t) over time is fitted with Kohlrausch−Williams−Watts (KWW) function: $r(t) = r(0) \cdot \exp(-(t/\tau_{seg})^{\beta})$, to extract the probe reorientation time (referred to here as the segmental relaxation time), $\tau_{seg}$, together with r(0), the anisotropy value at time zero, and β, the nonexponentiality factor. Note that, for samples without a large deformation, $I_{\perp,un}$ and $I_{\parallel,un}$ were almost same; while for melt-stretched samples or quenched samples at large strain ($\varepsilon_{eng}$ >50%), due to the alignment effect of the probe molecules along the deformation axis, the ratio between these two channels in unbleached area, g=$I_{\parallel,un}/I_{\perp,un}$ , can reach about 0.85; this has been taken into account in the calculation of r(t) by multiplying the fluorescence intensity of the perpendicular channel with factor g[34].

## RESULTS

In this section, results will be presented to show how the extent of strain-hardening influences the segmental dynamics, especially the relaxation time ($\tau_{seg}$)-strain rate ($\dot{\varepsilon}$) relationship, of PMMA glasses with and without melt-stretching, referred to as “melt-stretched” and “quenched” samples, respectively. It is useful to clarify the definitions of the mechanical parameters that will be utilized below: 1) All the stress data will be presented as “true stress”, which is determined from the instantaneous cross-sectional area of the sample rather than the initial value; 2) The strain values in these inhomogeneous tensile deformations will be local strains and two metrics will be used: local engineering strain and local true strain. Correspondingly, the strain rates presented will be local strain rates that were determined from the time derivatives of the strain profiles.

The mechanical behavior of quenched and melt-stretched PMMA glasses during constant global engineering strain rate deformation is illustrated in Fig. 1. Glasses prepared with these two protocols showed an identical stress response in the pre-yield regime, and both entered the strain-hardening regime immediately after yield. As expected[32] in the post-yield regime, the melt-stretched sample displayed a more prominent stress increase than the quenched sample. Although the strain rates were slightly different in the two deformations shown in Fig. 1, this does not account for difference in the post-yield moduli, as shown in Fig. S1 (see Supporting Information).

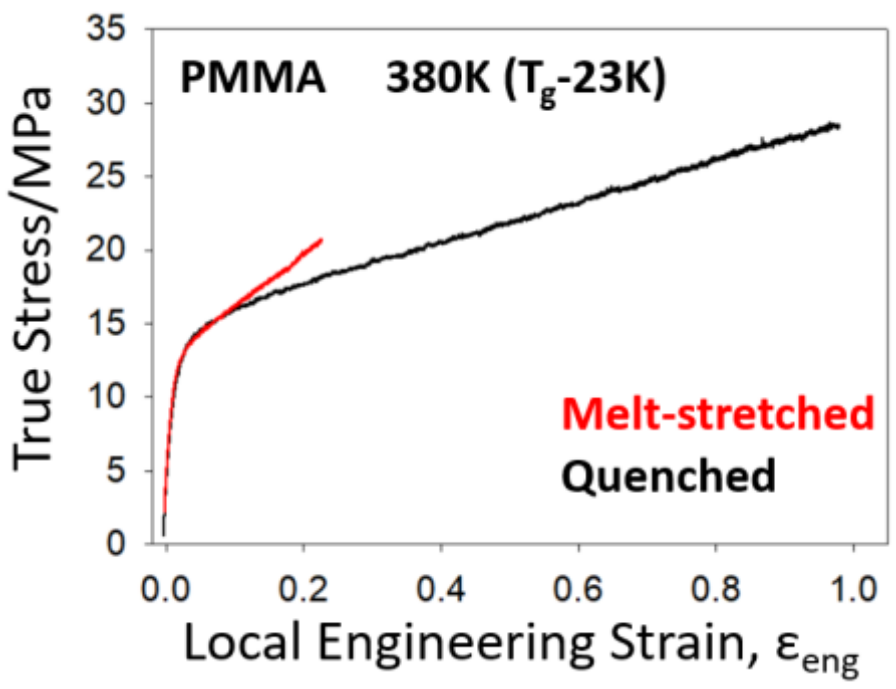

*Figure 1. Mechanical behavior of quenched and melt-stretched PMMA glasses under constant global engineering strain rate deformation. The strain rates are $2.7\times10^{-5}s^{-1}$ for the quenched sample and $3.1\times10^{-5}s^{-1}$ for the melt-stretched sample. The melt-stretched sample showed more prominent strain-hardening than the quenched sample.*

Fig. 2 shows the evolution of segmental dynamics and strain rates when a quenched sample was deformed deeply into the strain-hardening regime at $T_g$- 23K. Representative anisotropy decay curves are plotted in panel (a). In agreement with previous work[16], the decay became faster as the sample was deformed into the post-yield regime, indicating accelerated segmental dynamics. We observed that the β value from KWW fittings increased from 0.31 in the quiescent state to about 0.5 and stabilized at that value (see Fig. S2). Panel (b) shows the evolution of the relaxation time ($\tau_{seg}$), the engineering strain rate ($\dot{\varepsilon}_{eng}$), and the true strain rates ( $\dot{\varepsilon}_{true}$) from the start of the deformation until the experiment was stopped at a local engineering strain of about 100%. The log($\tau_{seg}$/s) values decreased by about 0.5 decade, from 2.8 to 2.3, as the sample reached a strain of 20%, and then increased after 70% strain. (Throughout this paper, base 10 log units are utilized.) Simultaneously, the local strain rates evolved in a reverse trend: the engineering strain rate rose from $10^{-4.5}$s to $10^{-4.2}$s at first and then decreased to about $10^{-4.4}$s, and the true strain rate increased by about 0.1 decade to $10^{-4.4}$s at the beginning and fell to $10^{-4.7}$s in the end. Results that are qualitatively similar to those shown in Fig. 2 were obtained for the melt-stretched samples.

As the most important result of this work, the interplay between the extent of strain-hardening, strain rates and segmental dynamics can be more clearly observed by plotting $\tau_{seg}$ vs. strain rates in log-log scale, as shown in Fig. 3, in which the engineering strain rate is plotted on the x-axis, and Fig. 4, in which the true strain rate is used instead. To make these plots, the post-yield data like that shown in Fig. 2 was binned across engineering strain ranges of about 0.1. Data from $15\%<\varepsilon_{eng}<30\%$ was collected in a single bin and is referred below to as the "early post-yield regime"; data for $\varepsilon_{eng}>30\%$ was collected in four bins and is described below as the "deep strain-hardening regime". The choice of local engineering strain rate for the abscissa of Fig. 3 is convenient but arbitrary. We expect, for deformations that enter the deep strain-hardening regime, true strain rate is the more fundamental parameter, and this is shown in Fig. 4. We want to emphasize that, although the global strain rate was held constant in these experiments, due to the inhomogeneous nature of these tensile deformations, neither the local engineering strain rate nor the local true strain rate were constant during deformation.

(a)

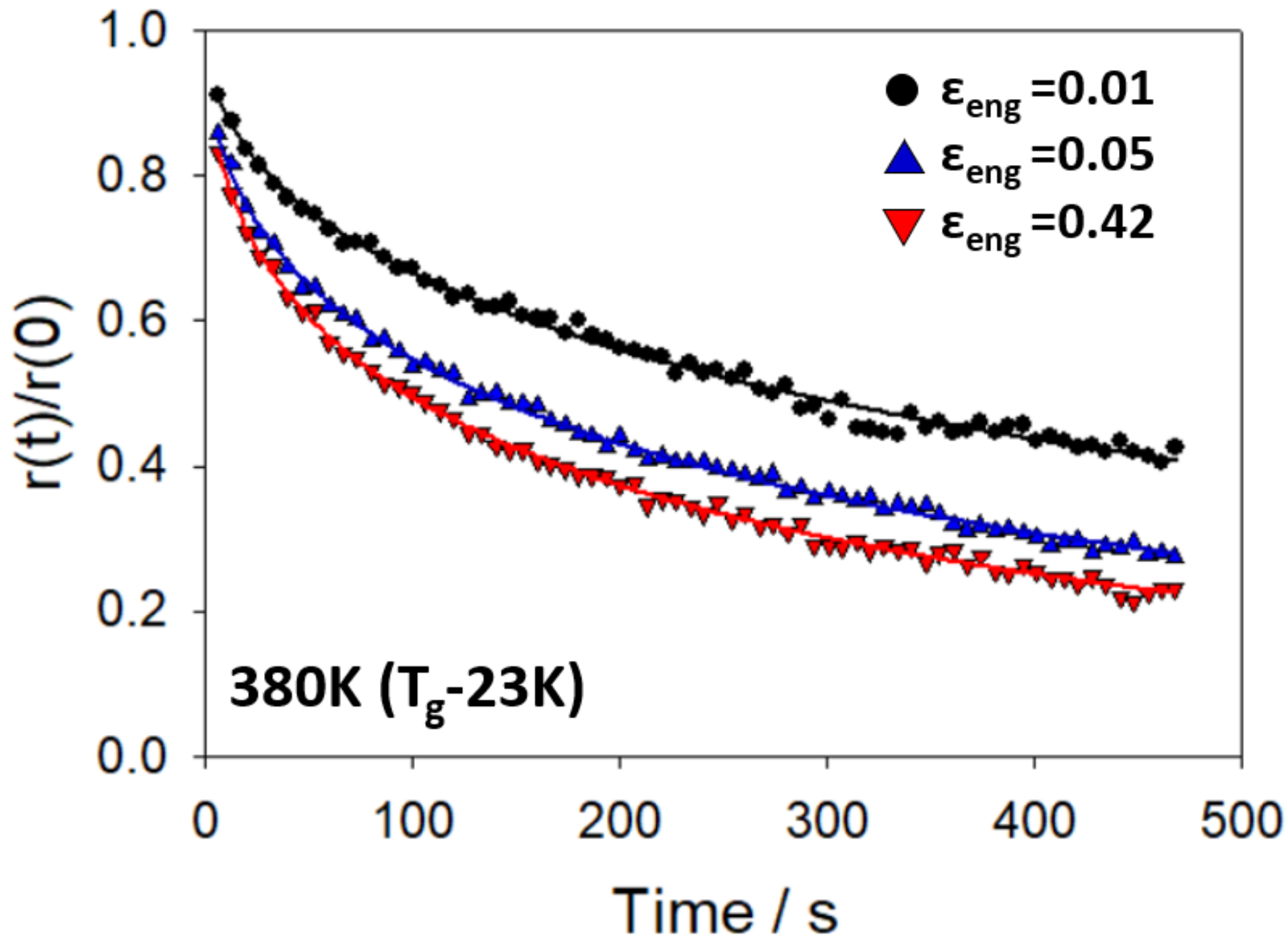

(b)

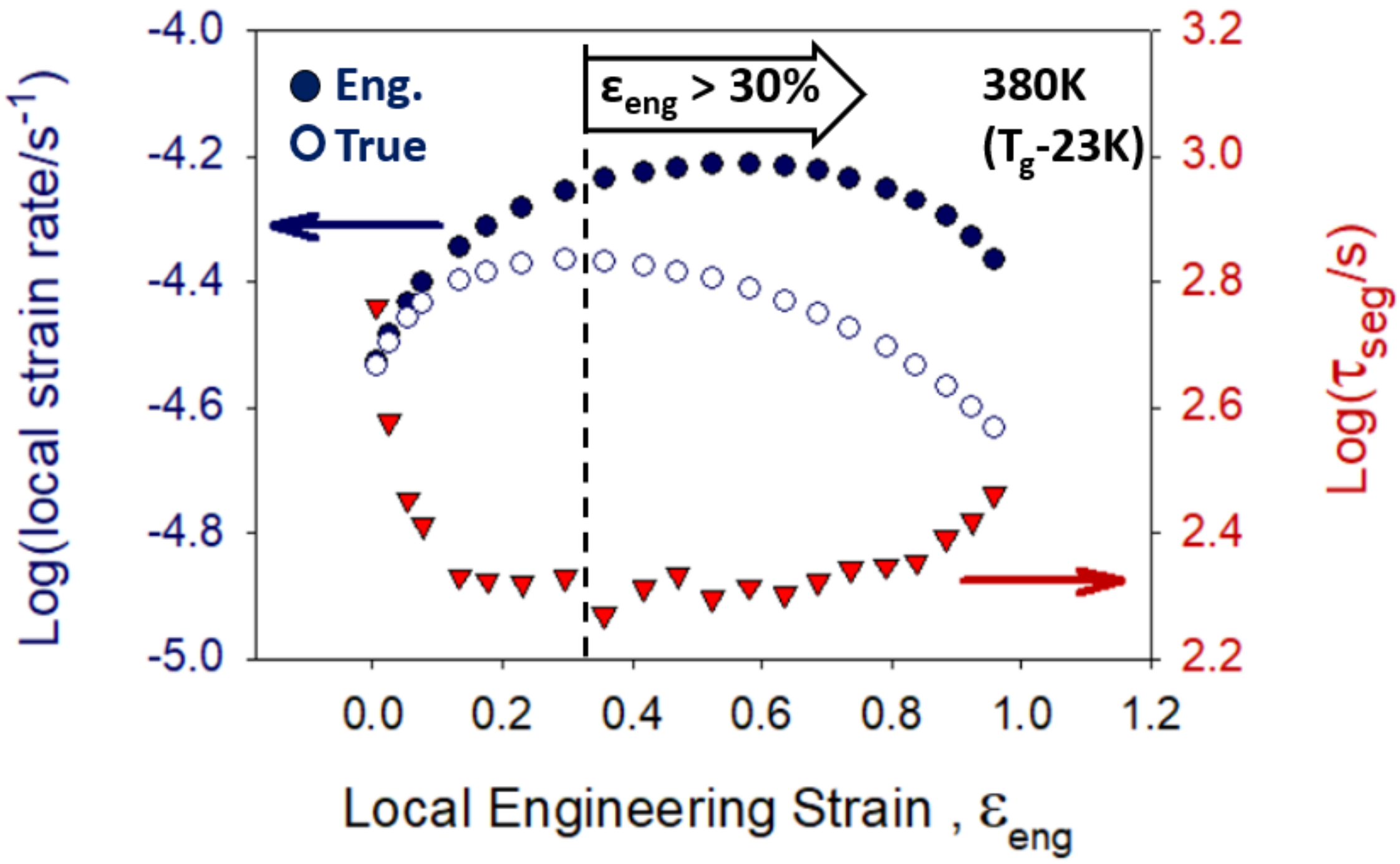

*Figure 2. The evolution of segmental dynamics and local strain rates for a quenched PMMA glass deformed at 380 K under constant global engineering strain rate of $3.1\times10^{-5}s^{-1}$. (a)The normalized anisotropy decay curves for photobleaching measurements performed at different local strains, indicating that the segmental dynamics accelerated as the sample was deformed into the post-yield regime.(b)The evolution of the local engineering/true strain rates and the relaxation time throughout the deformation. After reaching plateau values after yield, the strain rates decreased and the relaxation time increased in the strain-hardening regime.*

**Role of engineering strain rate.** In Fig. 3, we compare the relaxation time/ engineering strain rate relationship for two strain regimes, two types of sample preparation methods and at two temperatures. First, we look at the data from quenched samples at 370K ($T_g$ – 33K), which are represented by blue circles. Here, log($\tau_{seg}$/s) vs. log($\dot{\varepsilon}_{eng}$/s$^{-1}$) data measured at both the early post-yield regime (solid circles) and the deep strain-hardening regime (open circles) can be well described with a single line. To test the strength of this conclusion, we fitted the data in the early post-yield regime only and found essentially the same line, as shown in Fig. S3.

A similar (but not identical) relationship between log($\tau_{seg}$/s) and log($\dot{\varepsilon}_{eng}$/s$^{-1}$) is observed for melt-stretched samples at 370 K. Here in Fig. 3, as the strain range and the strain rate are controlled to be identical between the melt-stretched samples and the quenched samples in the early post-yield regime, the relationship between the extent of strain-hardening and segmental dynamics can be more clearly revealed. The data of melt-stretched samples (triangles) were fitted with dashed lines. In the figure, the melt-stretched data at 370 K is shifted downward by about 0.05 decade (12%) from the results of quenched samples. We attribute this small effect to the stronger strain-hardening in the melt-stretched samples, as shown in Fig. 1.

When we look at data in Fig. 3 acquired at 380K ($T_g$ -23K), similar features are observed. Again results from both the early post-yield regime (orange solid circles) and the deep strain-hardening regime (orange open circles) can be well described with a single line. At this temperature, within the noise of the measurements, there is not a significant difference between the quenched and the melt-stretched samples.

The slopes of the fitted lines for quenched samples agree within error with previous work on PMMA glasses deformed into the early post-yield regime[16]; here we observe slopes of -0.84±0.03 for $T_g$-33K and -0.77±0.06 for $T_g$-23K. The fitted slopes for the melt-stretched samples are -1.02±0.03 for $T_g$-33K and -0.87±0.15 for $T_g$-23K.

As a summary of Fig. 3, we find that, for a given engineering strain rate, the segmental dynamics are essentially the same for quenched samples deformed slightly past yield or deep into the strain-hardening regime. In addition, slightly faster segmental dynamics were observed for the melt-stretched samples.

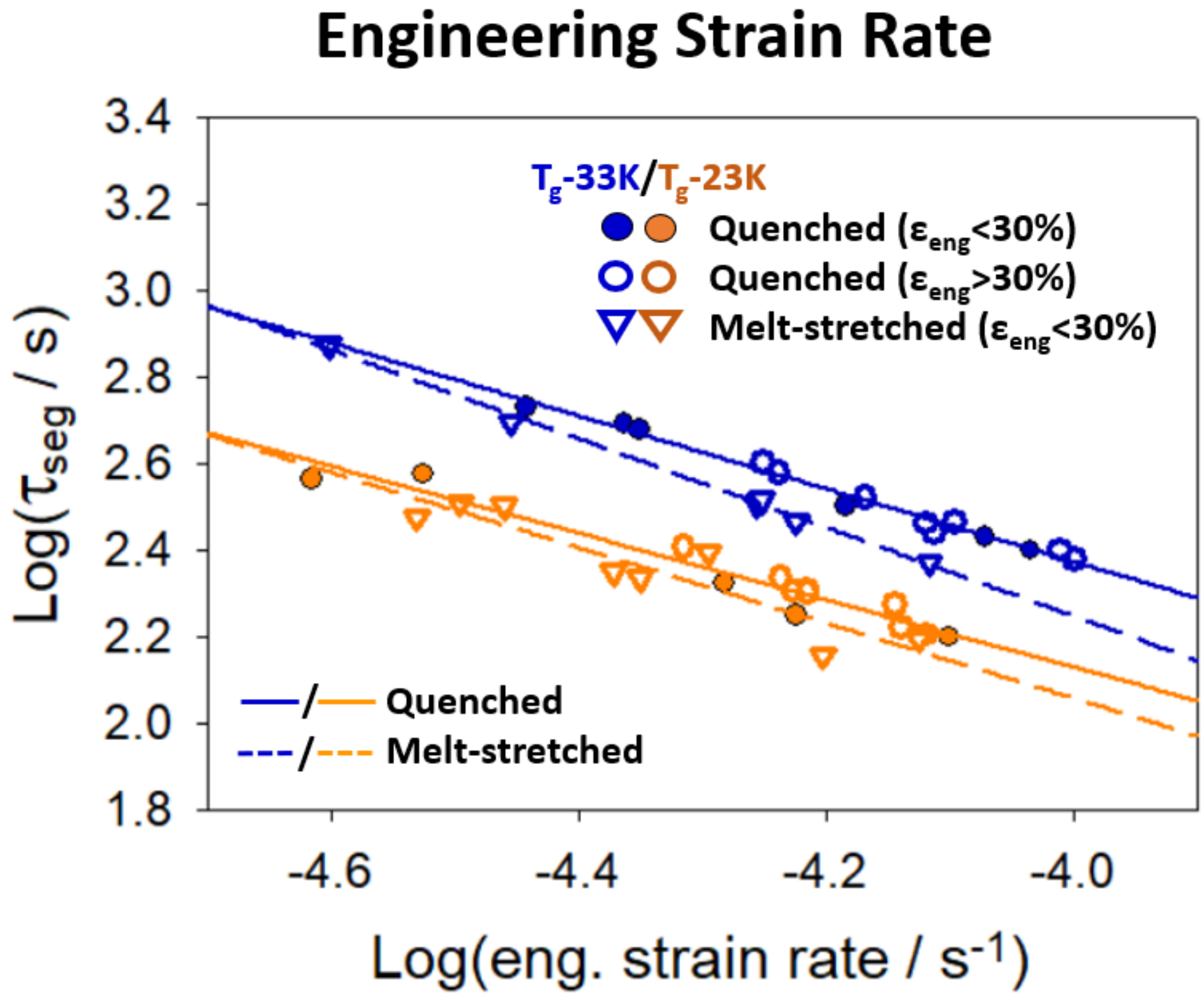

*Figure 3. Segmental relaxation times as a function of local engineering strain rate for the quenched and the melt-stretched PMMA glasses in the post-yield regime. Engineering strain rate is correlated closely with segmental dynamics even when the engineering strain approaches 100%. For both types of samples, the results are well described by linear fitting. The solid lines are fittings to the data of quenched samples, including data for* $\varepsilon_{eng}$ *up to 100%. At a given*

*engineering strain rate, the greater strain-hardening of the melt-stretched sample results in slightly faster segmental dynamics.*

**Role of true strain rate.** Fig. 4 shows that, at a given true strain rate, PMMA glasses deformed into the deep strain-hardening regime have faster segmental dynamics than the glasses in the early post-yield regime. To aid in understanding the figure, we note that the true strain rate is calculated from the instantaneous sample length, which increases as a deformation proceeds. Thus the true strain rate is smaller than the engineering strain rate during a tensile deformation and the difference between them increases as the strain becomes larger. As a result, when the engineering strain rate in Fig. 3 is replaced by the true strain rate here in Fig. 4, the data points from the deep strain-hardening regime are shifted to the left relative to the early post-yield regime data.

In Fig. 4, for each temperature, two deformations that entered the deep strain-hardening regime are illustrated and these are represented by different open symbols. In the deep strain-hardening regime, the true strain rate decreased during each deformation, so in the figure each deformation proceeds from the right of the figure to the left, as exemplified by a chain of black arrows connecting the data points for one deformation performed at $T_g$-23K. As the strain increases, the downward deviation of the relaxation times relative to the results in the early post-yield regime increases, up to a maximum of 0.15 decade in these experiments. This acceleration will be discussed further in the **Discussion** section, in comparison to the simulation results.

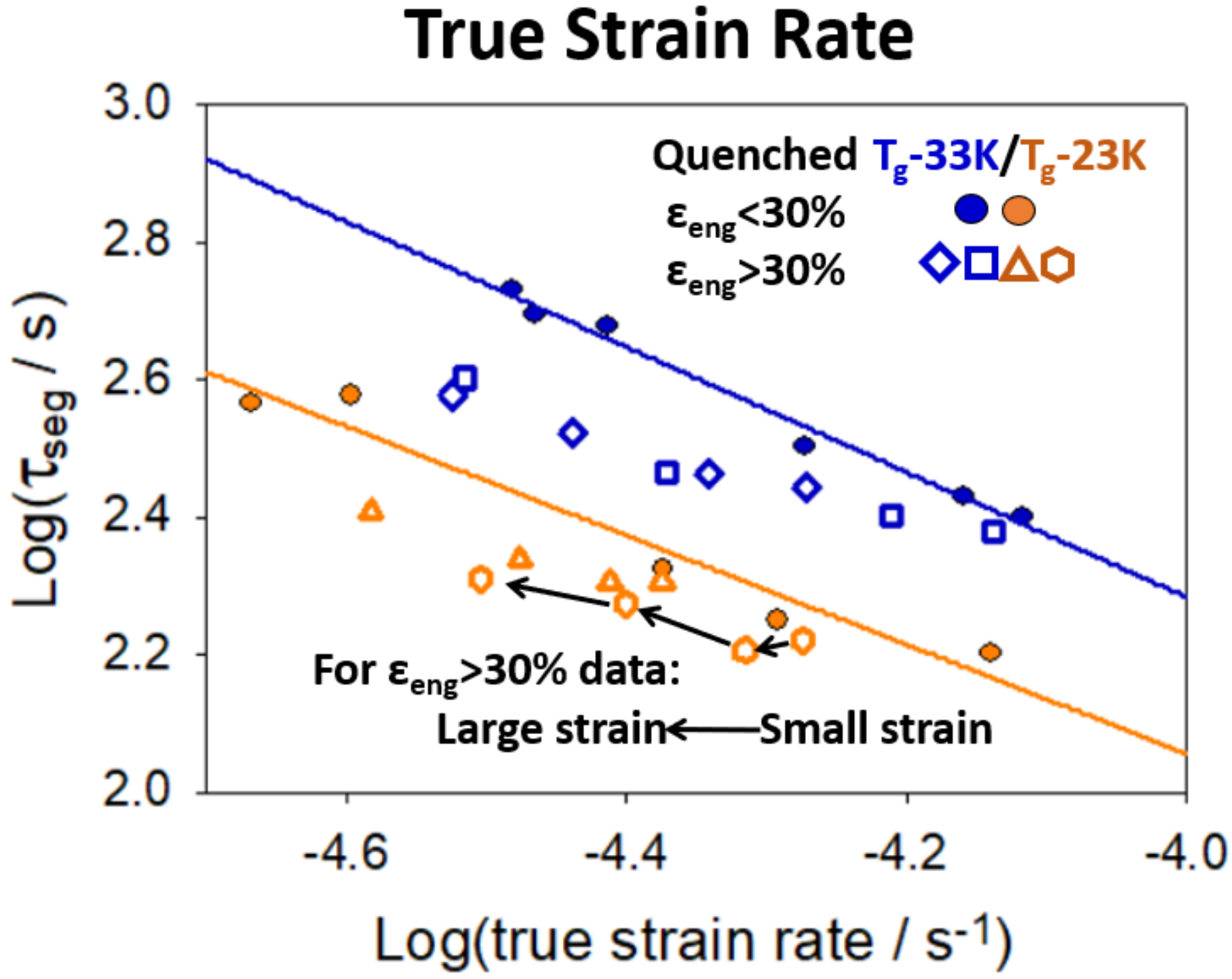

*Figure 4. Segmental relaxation times for the quenched PMMA glasses in the early post-yield regime ($\varepsilon_{eng}$<30%) and the deep strain-hardening regime ($\varepsilon_{eng}$>30%) with respect to local true strain rate. Each type of open symbol represents one deformation that has reached $\varepsilon_{eng}$>30%, and the chain of black arrows illustrates that the strain increased from right to left for each deformation. Linear fittings for $\varepsilon_{eng}$<30% data have slopes of -0.91±0.05 for $T_g$-33K and -0.79±0.10 for $T_g$-23K. When compared at the same true strain rate, segmental dynamics are faster at larger strains.*

**DISCUSSION**

In this section, we will compare our data with simulation and theoretical results. To make such a comparison, we assume that the segmental dynamics do not depend on the precise history of the deformation protocol. That is, we assume that the local strain rates in our experiments are changing sufficiently slowly that the results shown in Figures 3 and 4 would also be obtained if the local strain rates could be held constant. With this view in mind, we start the discussion.

The simulations done by Hoy and Robbins[25-26] are probably the earliest work that observed a correlation between the local particle mobility and the extent of strain-hardening. In most earlier theories[17-20], instead of the local rearrangement between particles, the entropic stress that arises between the chain entanglements was considered to be the origin of the strain-hardening of polymer glasses. Hoy and Robbins[25-26] simulated a bead-spring model under constant true strain rate deformations, in which the total stress arising from the input work was separated into the thermal component ($\sigma_Q$) and the energetic component ($\sigma_U$). They found that the thermal component, which dominated the early to moderate strain-hardening regime, arose from the irreversibly dissipated heat that was caused by friction between beads, instead of the reversible heat in the entropic stress picture. They also found that the contribution from the energetic component, which would be related to the bond stretching, became significant at a stretching ratio of about 2. Besides, systems with chain lengths shorter than the entanglement length were also able to show strain-hardening, which is another evidence against the idea that strain-hardening is solely caused by the entanglement network.  In Fig.4 of Ref.25, Hoy and Robbins showed that the evolution of $\sigma_Q$ almost exactly followed the trend of the plastic rearrangement rate calculated from the rearrangement between neighboring beads, both of which showed a general increasing trend as the strain became larger. In Fig. 4 of our work, we find qualitative agreement with this result; at a fixed true strain rate, an acceleration of local mobility is observed during strain-hardening. Hoy and Robbins also observed more prominent strain-hardening and acceleration of rearrangement rates in a system with higher entanglement density. In Fig. 3 of our work, we achieve a stronger strain-hardening through melt-stretching and we also observe a similar correlation between the strain-hardening and the molecular mobility. Our melt-stretched samples have a higher strain-hardening modulus and thus more energy dissipation during strain-hardening than quenched samples, and the melt-stretched samples show slightly faster segmental dynamics at least at one temperature ($T_g$-33K).

An acceleration of particle mobility during strain-hardening was also observed in simulations done by Rottler[27]. In that work, the decay of the self-intermediate scattering function was used as the probe of local segmental dynamics. Bead-spring models were deformed under constant true strain rate and the dynamics gradually accelerated, with the effect reaching a factor of about two at 100% true strain. In Fig. 5, we compare our results with Ref. 27 as a function of true strain. Here, we used the results in Fig. 4, in which true strain rate was plotted on the x-axis, to calculate the downward shift of $\log(\tau_{seg}/s)$ for samples deformed into the deep strain-hardening regime ($\varepsilon_{eng}$>30%), relative to the data for the samples deformed below 30% $\varepsilon_{eng}$. We can see that the experimental data generally match with the simulation result, both of which show an acceleration of about 0.15 decade at 80% true strain.

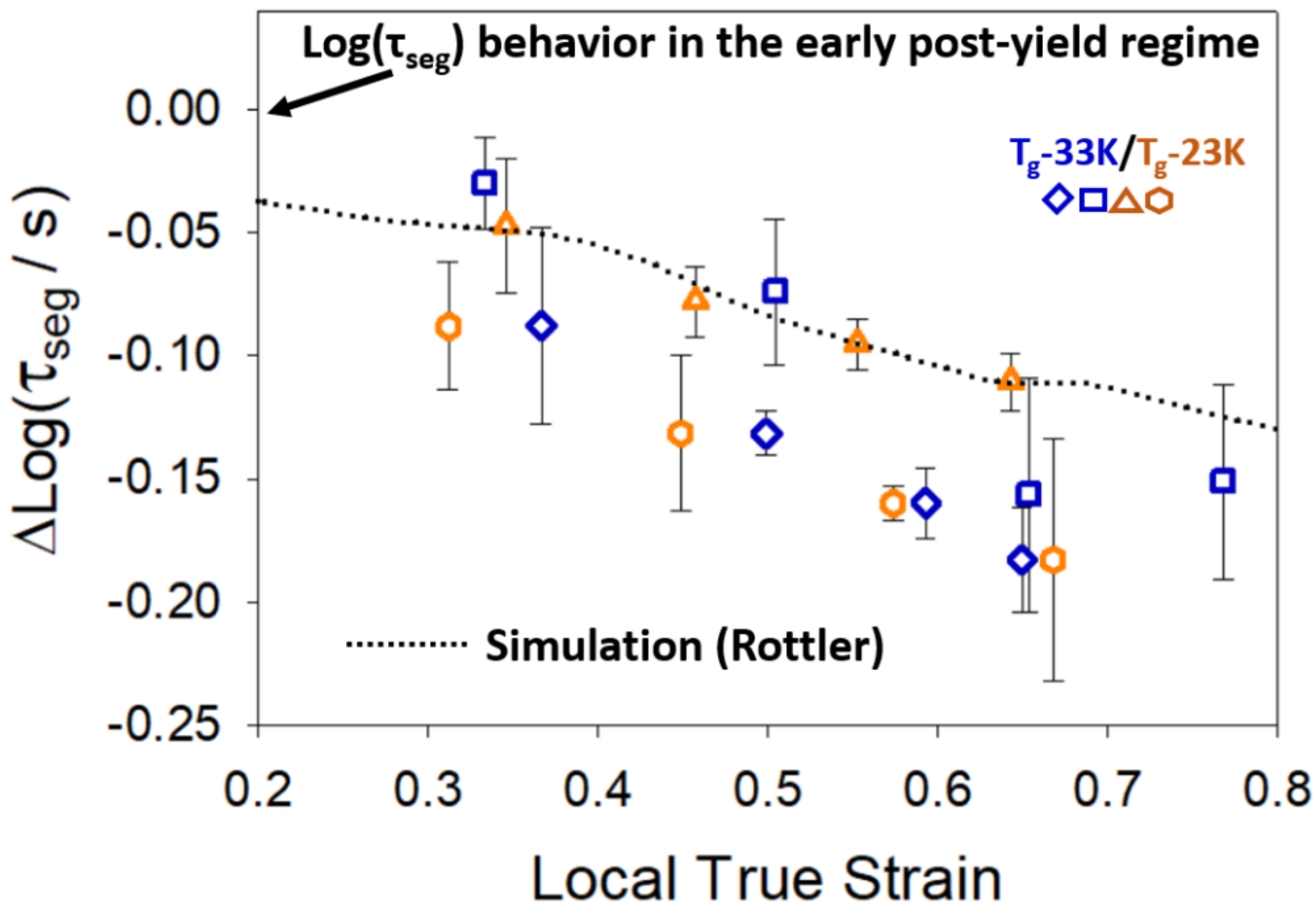

*Figure 5. Deviation of log($\tau_{seg}$/s) for PMMA glasses deformed into the deep strain-hardening regime ($\varepsilon_{eng}$>30%) relative to small strain data at the same true strain rate. The dotted line shows the simulation result from Rottler[27], which is the deviation of the relaxation time from the value in the early post-yield regime during a constant true strain rate deformation. The simulation and the experiments reported here show a comparable acceleration of segmental dynamics due to the strain-hardening, for the comparison at a fixed true strain rate. The error bars on the experimental points were determined from the standard deviations among measured relaxation time data.*

Atomistic simulations of polystyrene(PS) and polycarbonate(PC) performed by Vorselaars et.al[28] partially support the conclusions of the two simulation works above. With detailed structural information, the authors further observed that the strain-hardening was accompanied by an acceleration of the rate of nonaffine particle rearrangement, mostly resulted from conformational changes. And they also found that PC, which has stronger strain-hardening than PS, also showed larger increases of the rate of nonaffine particle rearrangement during deformation. In these simulations, the engineering strain rate was set to be constant, thus the acceleration of local mobility that they observed disagrees with what we measured: as shown in Fig. 3, samples in different strain regimes have identical segmental dynamics when compared at a given engineering strain rate. Apparently, their observation also disagrees with our interpretation of Rottler's simulation. The differences might result from the different simulated systems (atomistic vs. LJ), the different dynamic observables (the rate of nonaffine displacement vs. the decay rate of the self-intermediate scattering function), or other factors. The apparent disagreement between Ref. 28 and the experimental results might be caused by the different extents of chain networking: the simulations were performed on unentangled PC or PS while the experimental system was lightly-crosslinked PMMA. It's possible that the close correlation between the measured segmental dynamics and the engineering strain rate results from the ability of these PMMA glasses to retain the memory of their original states, as confirmed by the fact that these samples could recover to reproducible initial glassy states after annealing at a temperature above $T_g$ for long enough. Further simulations would be useful to understand this point.

Chen and Schweizer proposed a microscopic theory based upon the non-linear Langevin equation (NLE)[10, 29, 39] that relates strain-hardening with a change of segmental dynamics has also been proposed[29]. The model has shown the temperature and strain rate dependences of the magnitude of the strain-hardening that are consistent with the experimental and simulation results. And it predicts that the segmental dynamics decelerate during strain-hardening under a constant true strain rate. The picture described is that, during strain-hardening, the interchain excluded volume effect between anisotropic polymer chains increases, which suppresses local density fluctuations and slows down the segmental dynamics. Quantitatively, the segmental relaxation time at a true strain of 100% is predicted to increase by about a factor of two from the early post-yield regime value under a constant true strain rate of 0.001 $s^{-1}$. In contrast, the experimental work presented here and the simulations of Rottler[27] show a decrease in the segmental relaxation time during strain hardening (at a given true strain rate). We note that the theory of Chen and Schweizer did not include aging, mechanical rejuvenation, or collective elastic effects that were subsequently added to the NLE framework; it is possible that these modifications would qualitatively change the prediction of segmental relaxation times.

## CONCLUSION

The segmental dynamics of polymer glasses during strain-hardening are studied in this work, in which lightly-crosslinked PMMA glasses with or without melt-stretching were used as the sample systems. We found that, in agreement with previous simulations on bead-spring polymer melts, the segmental dynamics of the PMMA glasses without melt-stretching accelerate during strain-hardening under a constant true strain rate, an observation that contradicts the prediction of a recent theory. This finding supports the conclusion of computer simulations that strain-hardening originates from an elevated friction between polymer segments. In addition, we found that melt-stretched PMMA glasses, while having a stronger strain-hardening behavior, show slightly faster segmental dynamics than the samples without melt-stretching.

Melt-stretching is a powerful method for enabling the study of strain-hardening behavior on polymer glasses which would otherwise break before the strain-hardening regime is reached during tensile deformation. In the future, experiments might also be performed on linear (without crosslinking) PMMA glasses to study their segmental dynamics during strain-hardening. The comparison between the results from linear and crosslinked PMMA glasses can provide more insights on what effects the chain networking might have on the segmental dynamics during strain-hardening.

ASSOCIATED CONTENT

**Supporting Information**.

Strain-hardening moduli for the quenched and melt-stretched PMMA glasses, the evolution of the KWW β parameter during a deformation deep into the strain-hardening regime, the log($\tau_{seg}$/s) vs. log($\dot{\varepsilon}_{eng}$/s$^{-1}$) plot similar to Fig. 3 with the additional fitted lines that confirm the agreement between the results with different extents of strain-hardening, and the log($\tau_{seg}$/s) vs. log($\dot{\varepsilon}$/s$^{-1}$) plots similar to Fig. 3 and 4 with error bars added.

AUTHOR INFORMATION

**Corresponding Author**

M. D. Ediger - Department of Chemistry, University of Wisconsin-Madison, Madison, Wisconsin 53706, United States; orcid.org/0000-0003-4715-8473

Email: ediger@chem.wisc.edu

**Authors**

Enran Xing - Department of Chemistry, University of Wisconsin-Madison, Madison, Wisconsin 53706, United States; orcid.org/0000-0002-9609-6910

Trevor Bennin - Department of Chemistry, University of Wisconsin-Madison, Madison, Wisconsin 53706, United States; orcid.org/0000-0001-8509-6344

Masoud Razavi - Department of Chemistry, University of Wisconsin-Madison, Madison, Wisconsin 53706, United States; orcid.org/ 0000-0001-5967-2459

**Notes**

The authors declare no competing financial interest.

ACKNOWLEDGEMENTS

We thank the National Science Foundation (DMR-2002959) for support of this work.

For Table of Contents Only:

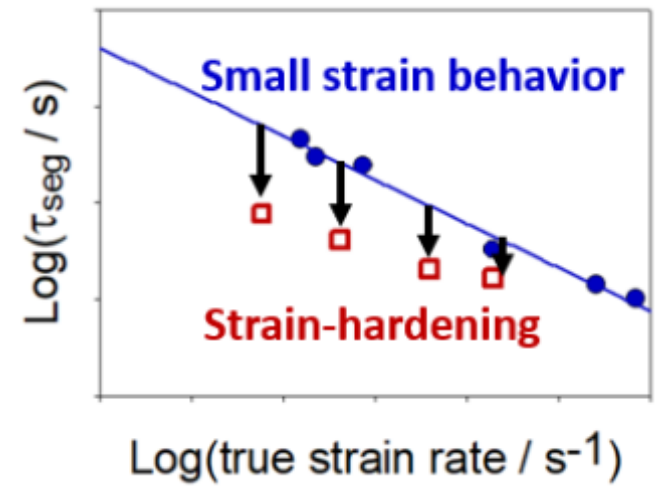

The editors may adjust the size of this figure as they see fit.

**SUPPLEMENTAL INFORMATION**

**Segmental Dynamics in the Strain-hardening Regime for Poly(methyl methacrylate) Glasses with and without Melt-stretching**

Authors: Enran Xing, Trevor Bennin, Masoud Razavi, M. D. Ediger*

Department of Chemistry, University of Wisconsin – Madison, Madison, Wisconsin 53706, United States

Corresponding author (ediger@chem.wisc.edu)

1. Strain-hardening moduli for the quenched and the melt-stretched PMMA glasses

In Fig. S1, we see that the strain-hardening modulus is largely independent of the strain rate for both types of PMMA glasses, for the range of strain rates utilized in this work. The strain-hardening modulus is calculated from the slopes the stress vs. Gaussian strain(g) curves, across ranges of $1.2<g<2.7$ for the quenched samples and $0.3<g<0.42$ for the melt-stretched samples. For reference, the Gaussian strain is defined as $g=\lambda^2-1/\lambda$, in which $\lambda$ is the stretching ratio.

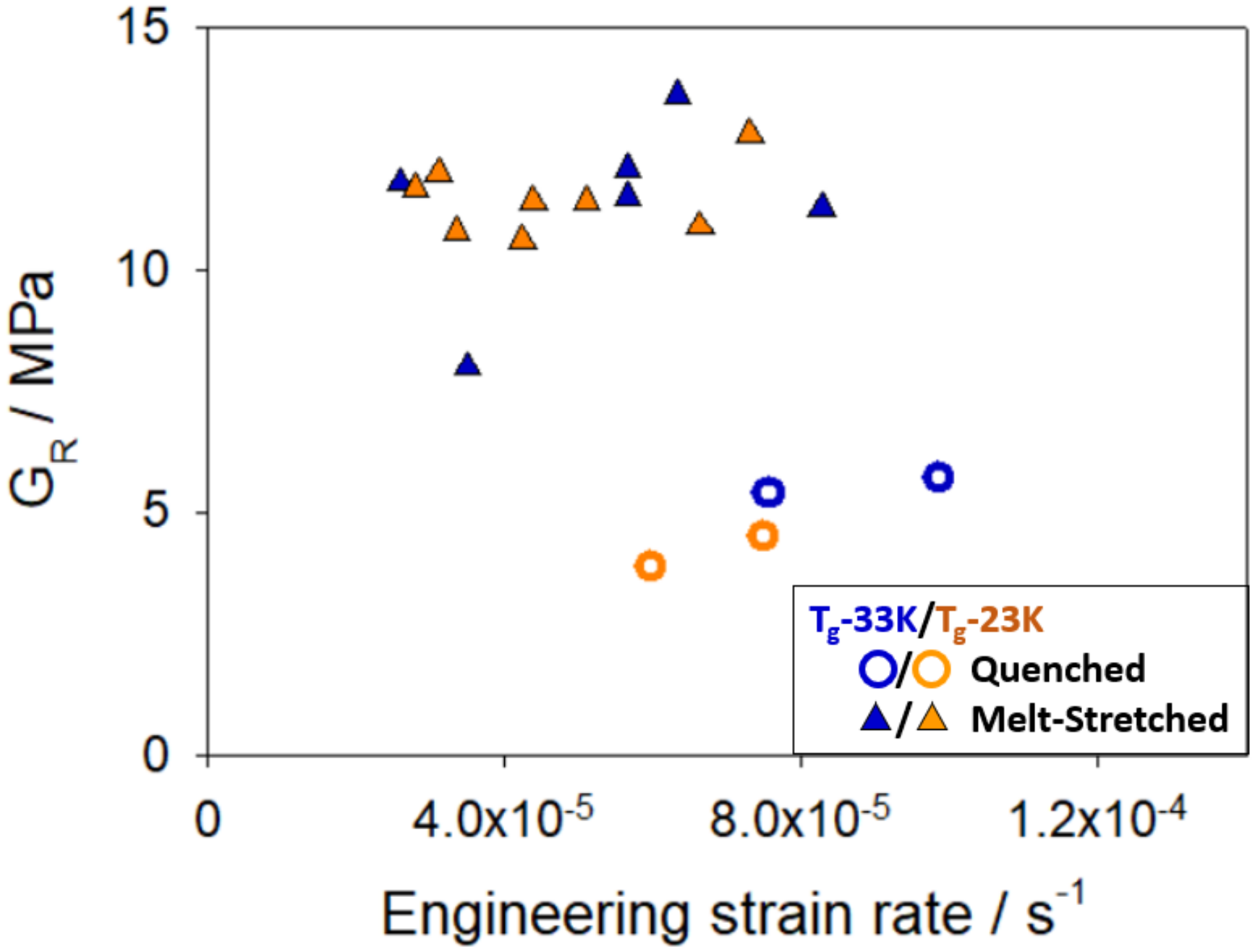

*Figure S1. The relationship between engineering strain rate and strain-hardening modulus.*

2. Evolution of the KWW β parameter during a deformation deep into the strain-hardening regime

Fig. S2 illustrates the β behavior during the deformation shown in Fig. 2. In support of the text in the main manuscript, we can see that β doesn't evolve significantly after the deformation begins. This result is typical for both the quenched and the melt-stretched samples.

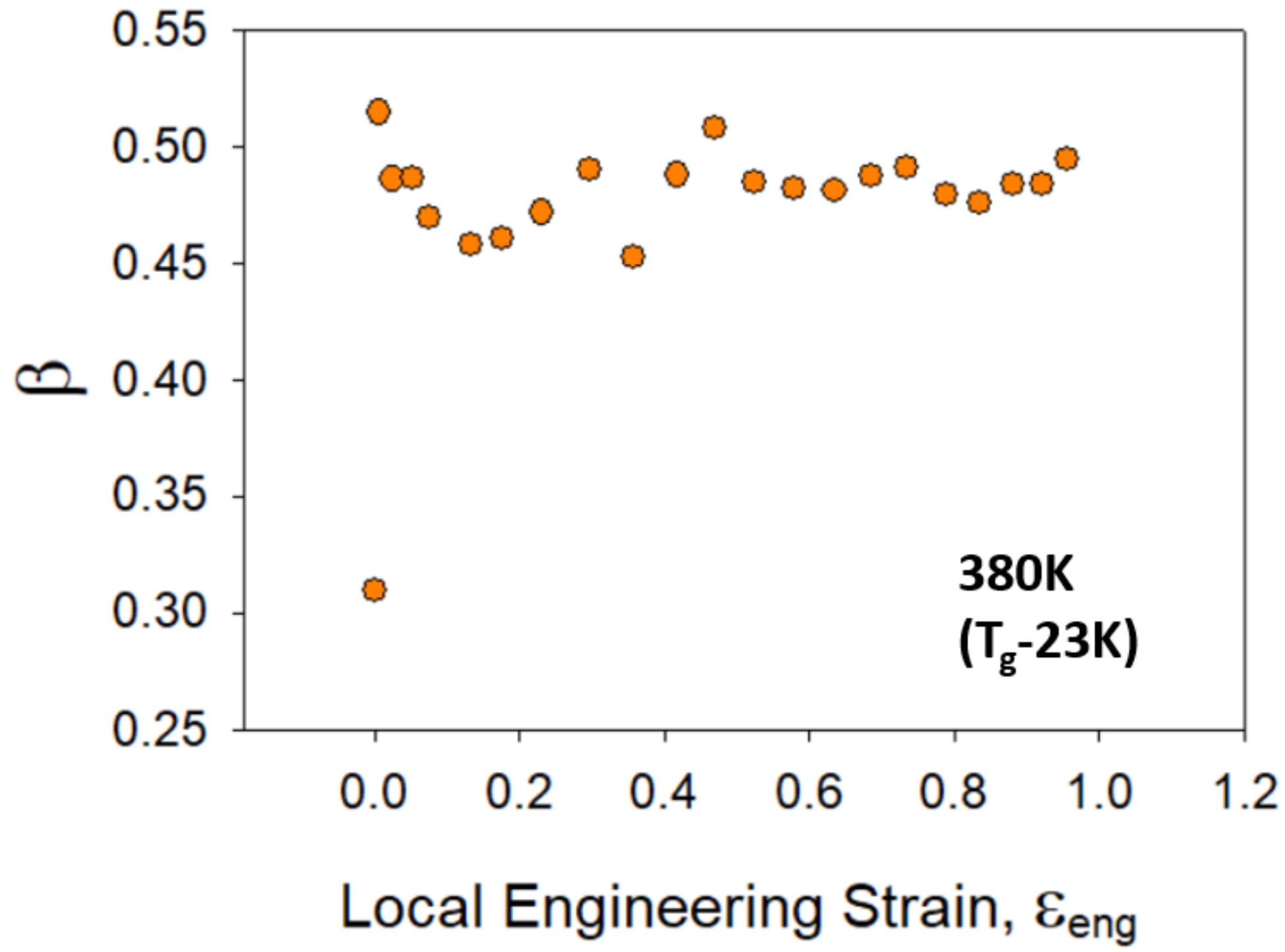

*Figure S2. The evolution of $\beta_{KWW}$ during the deformation of the quenched PMMA glass shown in Fig.2*

3. Role of engineering strain rate

In relation to Fig. 3 in the main text, the comparison between the fitted lines for quenched sample data in the early post-yield regime ($\varepsilon_{eng}$<30%) and over the entire strain range is shown in Fig. S3. These fitted lines almost overlap with each other for each of the testing temperatures, which supports the statement in the main text.

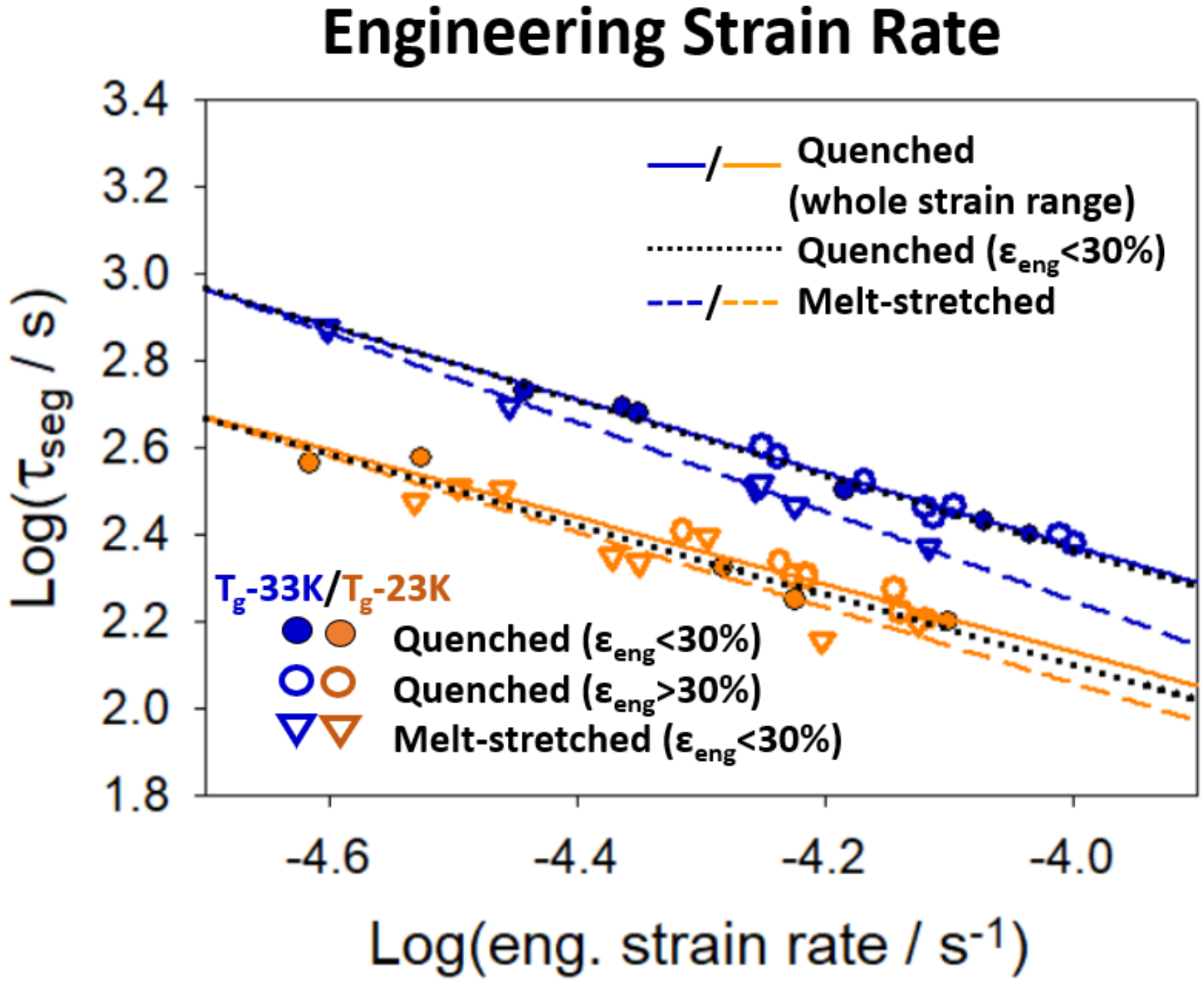

*Figure S3. Same as Fig. 3, but with the addition of linear fitted lines (black dotted lines) for the data of the quenched sample deformed below 30% local engineering strain, which are almost identical to the fitted lines for the data of the entire strain range.*

In Fig. S4, "error bars" calculated as the standard deviations of $\log(\dot{\varepsilon}_{eng}/s^{-1})$ and $\log(\tau_{seg}/s)$ are added to the data of Fig. 3. As both parameters were changing during the deformation, these bars show the range of strain rate and $\tau_{seg}$ that are binned together.

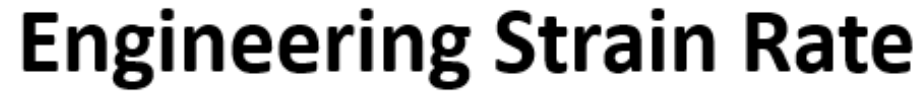

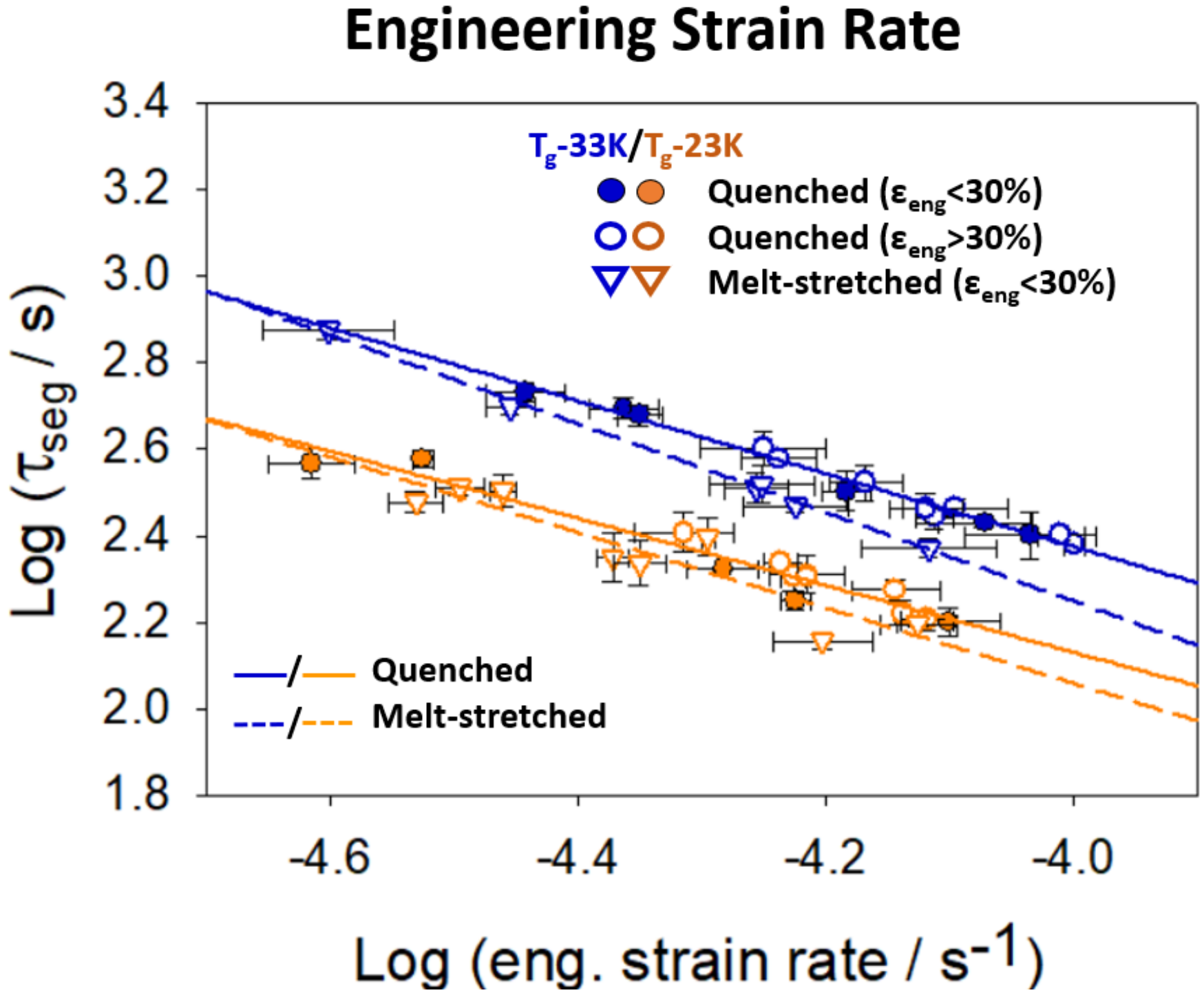

*Figure S4. Same as Fig. 3, with bars added that indicate the averaged ranges of $log(\dot{\varepsilon}_{eng}/s^{-1})$ and $log(\tau_{seg}/s)$.*

4. Role of true strain rate

In Fig. S5, similar to the case of the engineering strain rate, "error bars" estimated based on the standard deviations of $log(\dot{\varepsilon}_{true}/s^{-1})$ *and* $log(\tau_{seg}/s)$ are added to the data of Fig. 4.

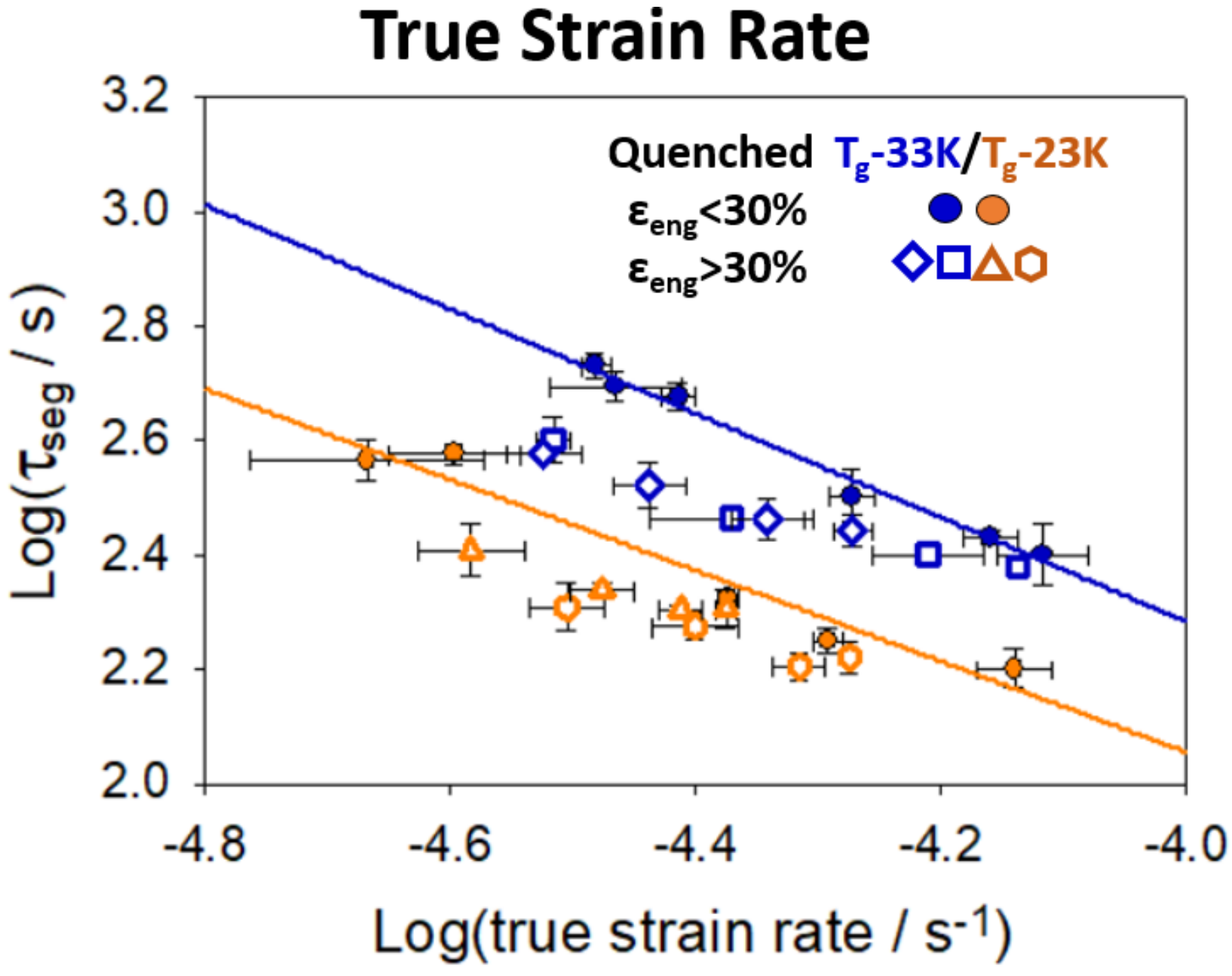

*Figure S5. Same as Fig. 4, with bars added indicating the averaged ranges of log(*$\dot{\varepsilon}_{true}$/s$^{-1}$*) and log(*$\tau_{seg}$*/s).*